\documentclass[a4paper,11pt]{article}
\usepackage{pos}

\title{Emergent geometric phase in time-dependent noncommutative quantum system}
\ShortTitle{Emergent geometric phase in noncommutative space-time}

\author*{Anwesha Chakraborty}

\affiliation{S. N. Bose National Centre for Basic Sciences\\
	Block-JD, Sector-III, Saltlake City, Kolkata-700106, West Bengal, India}

\emailAdd{chakra.an@gmail.com}

\abstract{Any effort to localise an event in the vicinity of the Planck length scale, only where the quantum
gravitational effects are predicted to be observed, will invariably result in gravitational collapse. One must postulate noncommutative (NC) algebra between space-time coordinates, which are
now elevated to the status of operators, in order to prevent such a situation from occurring.
On the other hand, a consistent formulation of Quantum mechanics itself, with time being an
operator is a challenging and longstanding problem. Here we have given a systematic way to
formulate non-relativistic quantum mechanics on 1+1 dimensional NC space-time (Moyal
type noncommutativity) in a user-friendly way, which mandates the formulation of an equivalent
commutative theory. Although the effect of noncommutativity of space-time should presumably
become significant at a very high energy scale, it is intriguing to speculate that there should be
some relics of the effects of quantum space-time even in a low-energy regime. With this motivation
in mind, we undertake the study of a time-dependent system, namely a forced harmonic oscillator
in NC space-time and have shown the emergence of a geometric phase, which vanishes if the
NC parameter is put to zero, proving the fact that, the occurrence of geometric phase is totally dependent on the non-commutativity of space-time.}

\FullConference{%
	Corfu Summer Institute 2022 "School and Workshops on Elementary Particle Physics and Gravity",\\
	28 August - 1 October, 2022\\
	Corfu, Greece}

\begin{document}
	\maketitle
	\section{Introduction}
	Over the past two decades, noncommutative geometry (NCG) has garnered a lot of theoretical interest in particle physics and condensed matter physics and is crucial to the understanding of quantum gravity models at the Planck scale. One can imagine a superposition of two mass distributions to explain the development of NC space-time or, more generally, quantum space-time. According to Penrose's argument in \cite{Roger}, this as a feedback through Einstein's GR, will result in a superposed geometry. Now, such a quantum space-time is likely to loose its time-translational symmetry resulting in the uncertainty of energy $\delta E$ and time $\delta t$, indicating a finite lifetime $\sim \frac{\hbar}{\delta E}$ of the system. This heuristic argument indicates that one needs to reformulate the quantum theory without classical time, rather time should be promoted to an operator-valued coordinate, along with other operator-valued spatial coordinates (see also \cite{fred,Dop}). The status of time in quantum gravity is an age-old problem. In fact, its status in quantum mechanics (QM) itself is a bit ambiguous. One can, in fact, recall Pauli's objection in this context \cite{pauli} and this ambiguity can result in other allied problems (for example see \cite{pull}). So one clearly needs to take the first step towards the formulation of a consistent version of Quantum Mechanics in quantum space-time first and eventually the QFT.  In this context, we can point out some earlier works in this direction \cite{szabo,apb,qft,liz} where some fascinating results were obtained, like discretization of time \cite{liz,time} corroborating similar observations made earlier by 't Hooft in the context of (2+1) dimensional quantum gravity \cite{hooft} and correction of spin-statistics theorem in NC spacetime leading to non-Pauli like transitions \cite{bal2}. Although the impacts of non-commutativity in the space-time sector should theoretically become substantial at (or before) extremely high energy scales, such as close to Planck scale energy, it is fascinating to imagine that some vestiges of the effects of non-commutativity in the low energy domain may exist \cite{sg1}, because of the inadequate decoupling mechanism between the high and low energy sectors. In  \cite{partha}, the authors have demonstrated a user-friendly method for formulating non-relativistic non-commutative Quantum Mechanics (NCQM) in 1+1 dimensional Moyal space-time, which necessitates the construction of an analogous commutative theory and also gets over the famous Pauli's argument \cite{pauli}. Here we use the formulation to further show the effect of noncommutativity in a time-independent system in the form of a geometric phase.\\ 
	The analysis of a typical time-independent system on NC \textit{space} results in new dynamics. The system's propagator's form is altered, and the wave function of the system also gets deformed. However, when such a system is placed in an NC \textit{space-time} background (i.e. one in which time also functions as an operator), no NC correction is obtained in the Hamiltonian or the spectrum of the system (see Appendix-A.1 of \cite{partha2}). This encourages us to investigate a \textit{time-dependent} system set up in an NC space time and search for any indications of noncommutativity, if any.  
 Thus our primary goal here is to look for the above mentioned signature, in the form of geometric phase in time dependent forced harmonic oscillator (FHO). For that we first need to  set up the formalism of QM itself of NC space time with time also being an operator which is the main content of section-2. Thereafter, in section-3, we discuss about the possibility of obtaining emergent Berry phase in a FHO system inhabiting NC space-time. In section-4 we conclude with some remarks and future directions.
	\section{Formulation of noncommutative quantum mechanics}
	The Hilbert-Schmidt (HS) operator based formulation of noncommutative QM was systematically devised, following \cite{nair1}, to formulate QM on `spatial' 2D noncommutative Moyal plane \cite{biswa,gauba}.  In this article we provide a brief review of our recent works on the formulation of QM on noncommutative (1+1)D Moyal space-time (based on the results in \cite{partha, partha2}) where we show how the HS operator formulation can be adapted to extract an effective, consistent and commutative quantum-mechanical theory. Now, before considering the quantum theory,  let's first talk about the advent of space-time bracket's noncommuting nature at the classical level itself. In light of this, take into account the following first-order form of a non-relativistic particle's Lagrangian in (1+1)D \cite{Deri}~:
	\begin{equation}
		L^{\tau,\theta}= p_{\mu}\dot{x}^{\mu}+\frac{\theta}{2}\epsilon^{\mu\nu}p_{\mu}\dot{p}_{\nu}-\sigma(\tau)(p_t+H),\,\,\,\,\,\mu ,\nu = 0,1\label{e73}
	\end{equation}
	where $x^{\mu}=(t,x)$ and $p_{\mu}=(p_t,p_x)$ are both considered as configuration space variables. The evolution parameter $\tau$ is bit arbitrary, except from the fact that it should be a monotonically increasing function of time $`t$'. Note, all the over-head dots in \eqref{e73} indicate $\tau-$ derivatives.
	On carrying out Dirac's analysis of constraints, one arrives at the following Dirac brackets between the phase space variables.
	\begin{equation}
		\{x^{\mu},x^{\nu}\}_D=\theta\epsilon^{\mu\nu};\,\,\,\,\,\{p_{\mu},p_{\nu}\}_D=0;\,\,\,\,\,\,\{x^{\mu},p_{\nu}\}_D=\delta^{\mu}\,_{\nu}\label{lev}
	\end{equation}
	Finally, the Lagrange multiplier $\sigma(\tau)$ enforces the following first class constraint in the system,
	\begin{equation}
		\Sigma= p_t+H \approx 0 \label{e87}
	\end{equation}
	and can be shown to generate the $\tau$ evolution of the system in the form of gauge transformation of the theory. 
	In order to begin the quantum mechanical analysis for this (1+1)D non-relativistic quantum mechanical system in the presence of the Moyal type space-time noncommutativity, we now raise the Dirac brackets in (\ref{lev}) to the level of commutator brackets:
	\begin{equation}
		[\hat{t},\hat{x}]=i\theta\label{e1} ~;~~~~~{\rm where,~}\theta~{\rm is~the~ NC ~parameter}
	\end{equation} 
	along with 
	\begin{equation}
		[\hat{p}_t,\hat{p}_x]=0,\,\,[\hat{t},\hat{p}_t]=i=[\hat{x},\hat{p}_x].\label{e2}
	\end{equation} 
	Note that we are working in the natural unit $\hbar= 1$ throughout this article.
	\subsection{Representation of the phase space algebra}
	A suitable representation of the NC coordinate algebra (\ref{e1}) can be shown to be furnished by the following Hilbert space.
	\begin{equation}
		\mathcal{H}_c=Span \left\{|n\rangle = \frac{(b^{\dagger})^n}{\sqrt{n!}}|0\rangle;\,\,b|0\rangle =\frac{\hat{t}+i\hat{x}}{\sqrt{2\theta}}|0\rangle =0 \right\}\label{e3}
	\end{equation} 
	We now introduce the associative NC operator algebra ($\hat{\mathcal{A}}_{\theta}$) generated by ($\hat{t},\hat{x}$) or equivalently by  ($\hat{b},\hat{b}^{\dagger}$)  acting on this configuration space $\mathcal{H}_c$ (\ref{e3}) as
	\begin{equation}
		\hat{\mathcal{A}}_{\theta}=\left\{|\psi)=\psi (\hat{t},\hat{x})=\psi(\hat{b},\hat{b}^{\dagger})= \sum_{m,n} c_{n,m}|m\rangle\langle n|\right\}\label{e59}
	\end{equation}
	which is the set of all polynomials in the quotient algebra ($\hat{\mathcal{A}}/\mathcal{N}$), subject to the identification of $[\hat{b},\hat{b}^{\dagger}]=1$. Thus, $\hat{\mathcal{A}}_{\theta}=\hat{\mathcal{A}}/\mathcal{N}$ is essentially identified as the universal enveloping algebra corresponding to (\ref{e1}), where $\hat{\mathcal{A}}$ is the free algebra generated by ($\hat{t},\hat{x}$) and $\mathcal{N}$ is the ideal generated by (\ref{e1}). This $\hat{\mathcal{A}}_{\theta}$ is not equipped with any inner product at this stage.\\ 
	We can now introduce a subspace $\mathcal{H}_q \subset \mathcal{B}(\mathcal{H}_c) \subset \hat{\mathcal{A}}_{\theta}$ as the space of `HS' operators, which are bounded and compact operators with finite HS norm $||.||_{HS}$, which acts on $\mathcal{H}_c$ (\ref{e3}), and is given by,
	\begin{equation}
		\mathcal{H}_q= \left\{\psi(\hat{t},\hat{x})\equiv \Big|\psi(\hat{t},\hat{x})\Big)\in \mathcal{B}(\mathcal{H}_c) ;\, \, ||\psi||_{HS}:=\sqrt{tr_c(\psi^{\dagger}\psi)} < \infty\right\} \subset \hat{\mathcal{A}}_{\theta}\label{e4}
	\end{equation}
	where $tr_c$ denotes trace over $\mathcal{
		H}_c$ and $\mathcal{B}(\mathcal{H}_{c})\subset \hat{\mathcal{A}}_{\theta}$ is a set of bounded operators on $\mathcal{H}_{c}$.  This space can be equipped with the inner product
	\begin{equation}
		\Big(\psi(\hat{t},\hat{x}),\phi(\hat{t},\hat{x})\Big):=tr_{c}\Big(\psi^{\dagger}(\hat{t},\hat{x})\phi(\hat{t},\hat{x})\Big)
		\label{iop}
	\end{equation}
	and therefore has the structure of  a Hilbert space on its own.  We use a different notation to denote the states of $\mathcal{H}_c$ and $\hat{\mathcal{A}}_{\theta}$  by $| .\rangle$ and $| . )$ respectively.  We now define the quantum space-time coordinates ($\hat{T},\hat{X}$) (which can be thought of as a representation of ($\hat{t}, \hat{x}$) and must be distinguished due to the fact that their domains of action are different, i.e., while $(\hat{T},\hat{X})$ act on $\mathcal{H}_q$, $(\hat{t},\hat{x})$ act on $\mathcal{H}_c$), as well as the corresponding momenta ($\hat{P_{t}},\hat{P}_{x}$)  by their actions on a state vector $|\psi(\hat{t},\hat{x})$ $\in \mathcal{H}_{q}$ as,\\
	\begin{align}
		&\hat{T}\Big|\psi(\hat{t},\hat{x})\Big)=\Big|\hat{t}\psi(\hat{t},\hat{x})\Big),\,\,\,\,\,\hat{X}\Big|\psi(\hat{t},\hat{x})\Big)=\Big|\hat{x}\psi(\hat{t},\hat{x})\Big),\nonumber\\
		&\hat{P}_x\Big|\psi(\hat{t},\hat{x})\Big)=-\frac{1}{\theta}\Big|[\hat{t},\psi(\hat{t},\hat{x})]\Big),\,\,\,\hat{P}_t\Big|\psi(\hat{t},\hat{x})\Big)=\frac{1}{\theta}\Big|[\hat{x},\psi(\hat{t},\hat{x})]\Big) \label{e5}
	\end{align}
	Thus, the momenta ($\hat{P_t},\hat{P_x}$) act adjointly and their actions are only defined in  $\mathcal{H}_{q}$ and not  $\mathcal{H}_c$. It may be easily verified now that ($\hat{T},\hat{X},\hat{P}_t,\hat{P}_x$) satisfies algebra isomorphic to the NC Heisenberg algebra just like \eqref{e2}.
	\subsection{Schr\"odinger equation and the \textit{physical} Hilbert space}
	Now it is evident that we cannot identify a counterpart to the common space-time eigenstate $|x,t\rangle $ in light of $\theta \neq 0$. However, by utilizing the coherent state, we are still able to reconstruct an efficient commutative theory. We select the Sudarshan-Glauber coherent state \eqref{e3} made up of the Fock states $|n\rangle$ that correspond to $\mathcal{H}_c$ as
	\begin{equation}
		|z\rangle = e^{-\bar{z}b+zb^{\dagger}}|0\rangle\,\in\mathcal{H}_c~;~~b|z\rangle=z|z\rangle\label{e43}
	\end{equation} 
	where $z$ is a dimensionless complex number and is given by,
	\begin{equation}
		z=\frac{t+ix}{\sqrt{2\theta}};\,\,\,\,\,\,\,\,\,t=\langle z|\hat{t}|z\rangle , x=\langle z|\hat{x}|z\rangle\label{A4}
	\end{equation}
	Note that $t$ and $x$ are not eigen values of the time and position operators rather they are expectation values of the respective operators in the coherent state basis (\ref{e43}) and will later be regarded as effective commutative coordinate variables. We can now construct the counterpart of coherent state basis in $\mathcal{H}_q$ (\ref{e4}), made out of the bases $|z\rangle \equiv |x,t\rangle$ (\ref{e43}), by taking their outer product as
	\begin{equation}
		|z,\bar{z})\equiv |z)=|z\rangle\langle z|=\sqrt{2\pi \theta}\,\,|x,t)\,\in \mathcal{H}_q \,\,\,\,\,\,\textrm{fulfilling}\,\,\,B|z)=z|z)\label{A8}
	\end{equation}
	where the annihilation operator $\hat{B}=\frac{\hat{T}+i\hat{X}}{\sqrt{2\theta}}$ is a representation of the operator $b$ in $\mathcal{H}_q$ (\ref{e4}). Note that the space-time uncertainty is saturated in this situation by $|z) \in \mathcal{H}_q$ (\ref{A8}), which suggests that such a state reflects a maximum localized \textquotedblleft{point}" or an event in space-time. Being a pure density matrix, this state can actually be thought of as an algebraic pure state. In fact this state being a pure density matrix can be regarded as a pure state of the algebra $\hat{\mathcal{A}}_{\theta}$ (\ref{e59}) and plays the role of a point, represented by Dirac's delta functional, in the corresponding commutative algebra $C^{\infty}(\mathbb{R}^2)$ describing (1+1) D commutative plane \cite{chaoba,anwe}. \\
	It can also be checked that the basis $|z,\bar{z})$ satisfies the over-completeness property:
	\begin{equation}
		\int \,\,\frac{d^2z}{\pi}|z,\bar{z})\,\star_{V}\,(z,\bar{z}|= \int dtdx \, |x,t) \star_{V} (x,t| = \textbf{1}_q,
		\label{vnc}
	\end{equation} 
	where $*_V$ represents the Voros star product and is given by,
	\begin{equation}
		\star_V=e^{\overleftarrow{\partial_z}\overrightarrow{\partial_{\bar{z}}}}=e^{\frac{i\theta}{2}(-i\delta_{ij}+\epsilon_{ij})\overleftarrow{\partial_i}\overrightarrow{\partial_j}};\,\,\,\,i,j=0,1;\,\,\,x^0=t, x^1=x
	\end{equation}
	Then the coherent state representation or the symbol of an abstract state $\psi(\hat{t},\hat{x})$ gives the usual coordinate representation of a state just like ordinary QM:
	\small
	\begin{equation}
		\psi(x,t)=\frac{1}{\sqrt{2\pi\theta}}\Big(z,\bar{z}\Big|\psi(\hat{x},\hat{t})\Big)=\frac{1}{\sqrt{2\pi\theta}}tr_{c}\Big[|z\rangle\langle z|\psi(\hat{x},\hat{t})\Big]=\frac{1}{\sqrt{2\pi\theta}}\langle z|\psi(\hat{x},\hat{t})|z\rangle \label{e45}
	\end{equation}
	\normalsize
	The corresponding representation of a composite operator say $\psi(\hat{x},\hat{t}) \phi(\hat{x},\hat{t})$ is  given by,
	\begin{equation}
		\Big(z\Big|\psi(\hat{x},\hat{t})\phi(\hat{x},\hat{t})\Big)=\Big(z\Big|\psi(\hat{x},\hat{t})\Big) \,\star_V\,\Big(z\Big|\phi(\hat{x},\hat{t})\Big).
		\label{como}
	\end{equation}
	This establishes an isomorphism between the space of HS operators $\mathcal{H}_q$ and the space of their respective symbols. Using (\ref{vnc})  the overlap of two arbitrary states ($|\psi),|\phi)$) in the quantum Hilbert space $\mathcal{H}_q$ can be written in the form
	\begin{equation} 
		(\psi|\phi) = \int dtdx ~ \psi^\ast(x,t) \star_{V} \phi(x,t)\label{innpro_voros}
	\end{equation}
	Therefore, to each element $|\psi(\hat{x},\hat{t})) \in \mathcal{H}_q$, the corresponding symbol is $\psi(x,t) \in L_{\star}^2(\mathbb{R}^2)$, where the $\star$-occurring in the subscript is a reminder to the fact that the corresponding norm has to be computed by employing the Voros star product. In order to obtain an effective commutative Schr\"odinger equation in coordinate space, we will introduce coordinate  representation of the phase space operators. To begin with, note that the coherent state representation of the actions of space-time operators  $\{\hat{X},\hat{T}\}$ on $|\psi)$ can be written by using (\ref{como}) as,
	\begin{equation}
		\Big( x,t\Big|\hat{X} \,\Big| \psi(\hat{x},\hat{t})\Big) =\frac{1}{\sqrt{2\pi\theta}}\Big(z,\bar{z}\Big|\hat{x}\psi\Big) = \frac{1}{\sqrt{2\pi\theta}} \, \left\langle z|\hat{x}|z\right\rangle \star_{V} (z,\bar{z}|\psi(\hat{x},\hat{t}))
	\end{equation}
	Finally on using (\ref{e45}), we have
	\begin{equation}
		\Big( x,t\Big|\hat{X}\Big| \psi(\hat{x},\hat{t})\Big) = X_\theta \, \Big(x,t\Big|\psi(\hat{x},\hat{t})\Big) = X_\theta \, \psi(x,t)~;~X_\theta = x+\frac{\theta}{2}(\partial_x-i\partial_t)\label{47}
	\end{equation}
	Proceeding exactly in the same way, we obtain the representation of $\hat{T}$ as
	\begin{equation}
		T_\theta = t+\frac{\theta}{2}(\partial_t+i\partial_x),\label{48}
	\end{equation}
	so that $[T_{\theta},X_{\theta}]=i\theta$ is trivially  satisfied. It is now trivial to prove the self-adjointness property of both $X_\theta$ and $T_\theta$, w.r.t. the inner product (\ref{innpro_voros}) in $\mathcal{H}_{q}$ by considering an arbitrary pair of different states $|\psi_1)$, $|\psi_2) \in \mathcal{H}_q$ and their associated symbols, just by exploiting associativity of Voros star product. Since momenta operators act adjointly, their coherent state representations are,
	\begin{equation}
		\Big(x,t\Big|\hat{P}_t \psi(\hat{x},\hat{t})\Big)= -i \partial_t\psi(x,t)~;~~ \Big(x,t\Big|\hat{P}_x \psi(\hat{x},\hat{t})\Big)=-i\partial_x\psi(x,t)
		\label{25}
	\end{equation}
	The effective commutative Schr\"odinger equation in NC space-time is then obtained by imposing the condition that the physical states $|\psi_{phy})=\psi_{phy}(\hat{x},\hat{t})$ are annihilated by the operatorial version of (\ref{e87}):
	\begin{equation}
		(\hat{P}_t+\hat{H})|\psi_{phy})=0;\qquad\psi_{phy}(\hat{x},\hat{t})\in \mathcal{H}_{phy}\subset \hat{\mathcal{A}}_{\theta} \label{e74}
	\end{equation} 
	where $\hat{H}=\frac{\hat{P}_x^2}{2m}+V(\hat{X},\hat{T})$. 
	We are now ready to write down the effective commutative time dependent Schr\"{o}dinger equation in quantum space-time by taking the representation of (\ref{e74}) in $|x,t)$ basis.
	Using (\ref{47},\ref{48},\ref{25}) we finally get,
	\begin{equation}
		i\partial_t \psi_{phy}(x,t)= \left[-\frac{1}{2m}\partial_x^2+ V(x,t)\, \star_{V}\right] \psi_{phy}(x,t)\label{e70}
	\end{equation}
	One can now obtain the continuity equation as,
	\begin{equation}
		\partial_t \rho_{\theta} +\partial_x J_{\theta}^{x}=0\label{e72}
	\end{equation}
	where 
	\begin{equation}
		\rho_{\theta}(x,t)= \psi_{phy}^*(x,t)\,\star_{V}\,\psi_{phy}(x,t)>0;\qquad J_{\theta}^{x}= \frac{1}{m}\mathfrak{Im}\bigg(\psi_{phy}^*\star_{V}(\partial_x \psi_{phy})\bigg) \label{prob}
	\end{equation}
Now $\rho_{\theta}(x,t)$ can be interpreted as the probability density at point $x$ at time $t$ given that it is positive definite. However, to achieve that and for a consistent QM formulation, we ought to have $\psi_{phy}(x,t)$ to be \textquotedblleft{well}-behaved" in the sense that it should be square integrable at a constant time slice:
	\begin{equation}
		\langle \psi_{phy} |\psi_{phy} \rangle_{\star\,t} =	\int_{-\infty}^{\infty} \,dx\,\, \psi_{phy}^*(x,t)\star_{V}\psi_{phy}(x,t) < \infty,
		\label{op}
	\end{equation}
	so that $\psi_{phy}(x,t)\in L_{\star}^2(\mathbb{R}^1)$  which is naturally distinct from $L_{\star}^2(\mathbb{R}^2)$. Equivalently , at the operator level, $\psi_{phy}(\hat{x},\hat{t})$ should belong to a suitable subspace of $\hat{\mathcal{A}}_{\theta}$ (\ref{e59}) which is distinct from $\mathcal{H}_q$, as the associated symbol for the latter is obtained from inner product defined for $L_{\star}^2(\mathbb{R}^2)$ (\ref{innpro_voros}). This is the main point of departure from the standard HS operator formulation of NCQM in (1+2)D Moyal plane with only spatial noncommutativity where time is treated as a c-parameter and one works with $\mathcal{H}_q$ or equivalently with a Hilbert space $L_{\star}^2(\mathbb{R}^2)$ for the corresponding symbols \cite{biswa,gauba}.
	\section{Forced harmonic oscillator and emergence of Berry phase}
	 We have used forced harmonic oscillator as a prototype system to look for any potential emergent geometric phases that could be signs of space-time noncommutativity. We therefore take the up the Hamiltonian of the forced harmonic oscillator in the following hermitian form for carrying out our analysis:
	\begin{equation}
		\hat{H}=\frac{\hat{P}_x^2}{2m}+\frac{1}{2}m\omega^2\hat{X}^2+\frac{1}{2}[f(\hat{T})\hat{X}+\hat{X}f(\hat{T})]+g(\hat{T})\hat{P}_x\label{e7}
	\end{equation}
	The corresponding effective commutative Schr\"odinger equation can be obtained by taking overlap of (\ref{e74}) in coherent state basis (\ref{A8},\ref{e45}),
	\begin{equation}
		i\partial_t \psi_{phy}(x,t)=\left[\frac{P_x^2}{2m}+\frac{1}{2}m\omega^2 X_{\theta}^2+\frac{1}{2}\{f(T_{\theta})X_{\theta}+X_{\theta}f(T_{\theta})\}+g(T_{\theta})P_x\right]\psi_{phy}(x,t)\label{e8}
	\end{equation}
	At this stage, it will be interesting to note that $X_{\theta}$ and $T_{\theta}$ can be related to commutative $x$ and $t$, defined in (\ref{A4}), by making use of similarity transformations,
	\begin{equation}
		X_{\theta}= SxS^{-1},~T_{\theta}=S^{\dagger}t(S^{\dagger})^{-1};~~S=e^{\frac{\theta}{4}(\partial_t^2+\partial_x^2)}e^{-\frac{i\theta}{2}\partial_t\partial_x}\label{M5}
	\end{equation}  
	This $S$, a non-unitary operator, can be used to define the following map,
	\begin{equation}
		S^{-1}\,:\,\,L_*^2(\mathbb{R}^1)\rightarrow L^2(\mathbb{R}^1) ~;~~i.e.~ S^{-1}\big(\psi_{phy}(x,t)\big):= \, \psi_c(x,t) \in L^2(\mathbb{R}^1) \label{M6}
	\end{equation}
	Now one can easily verify at this stage,
	\begin{equation}
		\Big \langle \psi_{phy}\, ,\, \phi_{phy} \Big\rangle_{\star,\,t}= \langle\psi_c\,\,,\phi_c\rangle _t\,\,\,\,\forall\,\psi_{phy},\phi_{phy}\,\in\, L^2_*(\mathbb{R}^1)\label{M7}
 	\end{equation}
	where we have made use of integration by parts and dropped some boundary terms. This equality shows that, we can replace non-local Voros star product,  with the local point-wise multiplication \textit{only} within the integral.   It should be emphasised here that, although the results of the integration as a whole are equal in both sides of (\ref{M7}), the integrands, by themselves are not : $\rho_{\theta}(x,t)=\psi_{phy}^*(x,t)\star \psi_{phy}(x,t)\,\ne\,\psi_c^*(x,t)\psi_c(x,t)=|\psi_c(x,t)|^2$. Thus here, one can't interpret $|\psi_c(x,t)|^2$ as the probability density at point $x$ at time $t$, unlike $(\psi_{phy}^*\star_{V}\psi_{phy})(x,t)$.  It is also important to keep in mind that the integrands themselves are components of two distinct algebras: non-commutative and commutative. These algebras are not clearly $\star$-isomorphic \cite{bal1} to one another, and the effect of non-commutativity is manifested through the considered dynamical model in a different way, as shown by the non-unitary transformation from $L_*^2(\mathbb{R}^1)$ to $L^2(\mathbb{R}^1)$ with help of the non-unitary transformation $S^{-1}$. \\
	Now, using (\ref{M5},\ref{M6}) in (\ref{e8}) and retaining terms upto linear in $\theta$, finally (\ref{e8}) can now be recast as,
	\begin{equation}
		i\partial_t \psi_c(x,t)=H_c\psi_c(x,t)\label{e13}
	\end{equation}
	where the corresponding effective commutative Hamiltonian $H_c$ is given by
	\begin{equation}
		H_c= \alpha(t) p_x^2+\beta x^2+\gamma(t)(xp_x+p_xx)+f(t)x+g(t)p_x=H_{GHO}+ f(t)x+g(t)p_x\label{H}
	\end{equation}
	Here $H_{GHO}$ stands for the Hamiltonian of a generalised time-dependent harmonic oscillator representing the first three terms. The last two terms represent perturbations linear in position and momentum in coordinate basis. And the various coefficients in (\ref{H}) are given by,
	\begin{equation}
		\alpha(t)=\frac{1}{2m}-\theta \dot{g}(t);\,\,\,\,
		\beta= \frac{1}{2}m\omega^2;\,\,\,\,
		\gamma(t)=-\frac{1}{2}\theta\dot{f}(t)\label{e14}
	\end{equation}
In order to diagonalize the whole Hamiltonian we try to diagonalize $H_{GHO}$ \cite{bdr} first. And for that we introduce annihilation and creation operators as follows:
\begin{equation}
	a(t)=A(t)[x+(B(t)+iC(t))p_x]\label{e15}
\end{equation}
One can also write down the corresponding creation operator $a^{\dagger}$ satisfying  $[a(t),a^{\dagger}(t)]=\textbf{1}$. The coefficients  $A, B , C$ and $\Omega$ are time dependent\footnote{The explicit forms of these functions can be found in \cite{partha2}}, as they depend upon $\alpha, \gamma$ and thus $\dot{f}(t)$ and $\dot{g}(t)$. Now note that  $\dot{f}(t)$ and $\dot{g}(t)$ are also periodic functions of \textquoteleft{$t$}' like $f(t)$ and $g(t)$. We just impose the condition that $\dot{f}(t)$ and $\dot{g}(t)$ are slowly varying function of time. This will facilitate the use of adiabatic approximation in the system under consideration. So we can say that $A, B , C$ are slowly varying function of time and we shall in future neglect the second and higher order time derivative of these variables in our adiabatic approximation.
With this, the Hamiltonian (\ref{H}) can be written in terms of the ladder operator as, 
\begin{equation}
	H_c= \Omega(t) \Big(a^{\dagger}(t)a(t)+\frac{1}{2}\Big)+P(t)a(t)+\bar{P}(t)a^{\dagger}(t)\label{H1}
\end{equation}
where, $$P(t)= A(t)[C(t)f(t)+i(B(t)f(t)-g(t))]=\sqrt{\frac{\beta}{\Omega(t)}}\left[\frac{\Omega}{2\beta}f(t)+i\left(\frac{\gamma(t)f(t)}{\beta}-g(t)\right)\right]$$
and $\Omega(t)$ can be identified with the \textquotedblleft{instantaneous} frequency". Although this $\Omega(t)$ can be considered to be a slowly changing function of time, compatible with the adiabatic nature, the other time dependent functions $P(t)$ and $\bar{P}(t)$ are somewhat unclear because they include phrases like the product of a fast varying and a slow varying function. Thus, we cannot claim that the whole Hamiltonian evolves slowly under temporal development. In order to circumvent this problem, let us try to find a suitable time dependent unitary transformation $\mathcal{U}(t)$, transforming the wave function $\psi_c(x,t)$ in (\ref{e13}) as,
\begin{equation}
	\psi_c(x,t)\,\to\, \tilde{\psi}_c(x,t)=\mathcal{U}(t) \psi_c(x,t);\,\,\,\,\,\mathcal{U}^{\dagger}(t)\mathcal{U}(t)=1\label{M9}
\end{equation}
With this, the corresponding Hamiltonian will transform under this time-dependent  unitary transformation as
\begin{equation}
	H_c\,\to\, \tilde{H}_c=\mathcal{U}(t)H_c\mathcal{U}^{\dagger}(t) - i\mathcal{U}(t)\partial_t \mathcal{U}^{\dagger}(t)\label{H3}
\end{equation}
This demonstrates that an effective Hamiltonian $\tilde{\mathcal{H}}_c$, which is obtained by supplementing $\mathcal{U} H_c
\mathcal{U}^{\dagger}$ by an appropriate \textquotedblleft{connection}" like term $-i\mathcal{U}\partial_t \mathcal{U}^{\dagger}$, now governs the time evolution of the transformed states $\tilde{\psi}_c(x,t)$.
\begin{equation}
	i\partial_t\tilde{\psi}_c(x,t)=\tilde{H}_c \tilde{\psi}_c(x,t) \label{H2}
\end{equation}
Indeed, there exists a $\mathcal{U}(t)$ given by $
	\mathcal{U}(t)= e^{-(wa-\bar{w}a^{\dagger}+il)}$, which can be identified with an element of non-abelian Heisenberg group $\mathfrak{h}$. $w$ and $l$ are some time-dependent complex and real functions \cite{partha2}. Now the Hamiltonian $\tilde{H}_c$ can be put into the diagonal form
	\begin{equation}
		\tilde{H}_c=\Omega(t)(a^{\dagger}a+\frac{1}{2})= H_{GHO}\label{H4}
	\end{equation}
	Carrying out the analysis in the Heisenberg picture where we make use of the adiabaticity of the parameters $\alpha$ and $\gamma$ by dropping their second and higher order time derivatives, we find after evolution through a cycle of time period $t=\mathcal{T}$ the creation operator takes the following form,
	\begin{equation}
		a^{\dagger}(\mathcal{T})=a^{\dagger}(0)exp \left[i \int _0^{\mathcal{T}} \Omega\,\, d\tau + i\int_0^{\mathcal{T}} \left(\frac{1}{\Omega}\right)\frac{d\gamma}{d\tau} d\tau \right]\label{e35}
	\end{equation}
	The second term in the exponential of (\ref{e35}) represents an additional phase  over and above the dynamical phase $e^{i\int \Omega(t)dt}$. This phase can now be written in a more familiar form of Berry phase \cite{berry}, given as a functional of the closed loop $\Gamma$,
	\begin{equation}
		\Phi_G[\Gamma] =\oint_{\Gamma=\partial S} \frac{1}{\Omega} \nabla_\textbf{R}\gamma . d\textbf{R}=-\frac{\theta}{2}\int\int_S \nabla_{\textbf{R}}\left(\frac{1}{\Omega}\right) \times \nabla_{\textbf{R}} \Big(\dot{f}(t)\Big)\, .\, d\textbf{S}
		\label{e36}
	\end{equation}
 We would like to make some pertinent observations before we conclude this section:
 \begin{itemize}
     \item{ The expression (\ref{e36}) explicitly depends on the NC parameter $\theta$ and disappears in the $\theta \to 0$ commutative limit.
	Hence, the non-commutativity of space-time is necessary for the emergence of the geometric phase in this context. In this respect, it should be noted that the authors in \cite{bdr} have demonstrated that a system must have the dialatation term in the Hamiltonian in order to produce the geometric phase. The forced harmonic oscillator in usual space-time, does not contain such term, but putting the system in an NC space-time automatically gave rise to a dialatation term in the effective commutative Hamiltonian (\ref{H}).}
 \item{Finally observe that, had we worked in the Schr\"{o}dinger's picture, we would have obtained the same geometric phase $\Phi_G[\Gamma]$ acquired by the wave function $\tilde{\psi}_c(x,t)$ (\ref{M9}) in time $\mathcal{T}$. Eventually since the original wave function $\psi_c(x,t)$ is simply $\mathcal{U}^{\dagger}(t)\tilde{\psi}_c(x,t)$ with $\mathcal{U}(t)$ being a linear and unitary operator, it is clear that $\psi_c(x,t)$ too will acquire the same Berry phase in time $\mathcal{T}$, as the Berry phase, being a simple number (\ref{e36}) will not be affected by action of the Heisenberg group.}
 \section{Remarks}
 It is difficult to experimentally detect Planck scale phenomena in a non-relativistic (NR) system that involve the quantum structure of space-time. Yet, researchers have attempted to develop convincing theories in some earlier studies in the literature \cite{cast,bek} in an effort to determine the implications of the quantum structure of space-time in the low energy regime. That inspired us to conduct this current work, which is to examine the fingerprints of Planck scale physics in the NR quantum system. Here, we'll sum up our research in a few sentences.\\
 We have demonstrated that a time-dependent forced harmonic oscillator system experiences a geometric phase shift when it is placed into a non-commutative space-time. It is demonstrated that the Hamiltonian of the effective commutative system is that of a time-dependent generalised harmonic oscillator. We then came up with the equation of motion for the ladder operators in the Heisenberg picture, which demonstrates that when a system is moved adiabatically around a closed loop, it can produce an additional phase in addition to the dynamical phase, depending on the geometry of the parameter space. This phase depends on the non-commutative parameter $\theta$ and vanishes in commutative limit $\theta \to 0$, proving its origin in the noncommutative nature of space-time.\\
  Finally, we would like to comment on some future prospects of our work. It was demonstrated in \cite{mehta} that a coherent state continues to be coherent at all times if the system Hamiltonian is that of a time-varying forced harmonic oscillator. We can extend this computation for our system, which is essentially FHO in NC space-time. The effective commutative system which is a generalised forced harmonic oscillator (GFHO), will create squeezed coherent states \cite{xu}. This motivates us to study the evolution of squeezed coherent states and its possible implications for quantum optics and quantum information theory.\\
Furthermore, note that the idea of noncommutativity in the space-time sector may also be introduced through the non-relativistic,
second quantization formalism. This would in fact be a good place to start when thinking about NC space-time because, unlike in QM, here, space and time can naturally be treated equally in the sense that they are simply c-number parameters. Crucially, the second quantization formalism enables us to do an analysis similar to NC quantum mechanical analysis in the first quantized formalism of the effects of space-time noncommutativity on the one-particle sector of quantum field theory \cite{nair}.
	\end{itemize}
 \section*
 {Acknowledgment}
The author thanks the organizers of the Corfu Summer Institute and
the Workshop on Noncommutative and generalized geometry in string theory, gauge theory and related physical models for their kind invitation.

\end{document}